\documentclass{article}
\usepackage{spconf,amsmath,graphicx,hyperref}

\title{Do we really need Self-Attention\\
for Streaming Automatic Speech Recognition?}
\name{Youness Dkhissi$^{1,2}$, Valentin Vielzeuf$^1$, Elys Allesiardo$^1$, Anthony Larcher$^2$}
\address{1. Orange Innovation \{firstname\}.\{lastname\}@orange.com\\
    2. LIUM, Le Mans Université Avenue Olivier Messiaen,72085 Le Mans CEDEX 9, France}

\usepackage{tikz}
\usepackage{multirow}
\usepackage{booktabs}
\usepackage{subcaption}

\usetikzlibrary{shapes, arrows.meta, positioning, fit}

\begin{document}
\ninept
\maketitle
\begin{abstract}
Transformer-based architectures are the most used architectures in many deep learning fields like Natural Language Processing, Computer Vision or Speech processing. It may encourage the direct use of Transformers in the constrained tasks,
without questioning whether it will yield the same benefits as in standard tasks.

Given specific constraints, it is essential to evaluate the relevance of transformer models. This work questions the suitability of transformers for specific domains.
We argue that the high computational requirements and latency issues associated with these models do not align well with streaming applications.
Our study promotes the search for alternative strategies to improve efficiency without sacrificing performance.

In light of this observation, our paper critically examines the usefulness of transformer architecture in such constrained environments. As a first attempt, we show that the computational cost for Streaming Automatic Speech Recognition (ASR)  can be reduced using deformable convolution instead of Self-Attention. 
Furthermore, we show that Self-Attention mechanisms can be entirely removed and not replaced, without observing significant degradation in the Word Error Rate.
\end{abstract}
\section{Introduction}
Transformers\cite{vaswani2017attention} have become the \textit{de facto} architecture for Natural Language Processing (NLP). Most NLP solutions build on top of a model pre-trained on a large dataset to take advantage of Self-Attention mechanisms and capture the long-range dependencies between the input sequence elements.
After this initial pre-training, the model can be fine-tuned on the downstream tasks.
The gain of performance brought by Transformers has encouraged the use of this architecture and its adaptation for other fields. 

Vision Transformer\cite{dosovitskiy2021an} (ViT) have ended the dominance of Convolutional Neural Network (CNN) based architectures in the State Of The Art (SOTA) of many Computer Vision tasks. 
Also, in the Speech field, the integration of the Self-Attention mechanism has been very effective in improving the SOTA on different tasks. 
In particular, for Automatic Speech Recognition (ASR), Conformers\cite{gulati2020conformer} have become the SOTA architecture thanks to the combination of Self-Attention modules that capture long-range dependencies and Convolutional modules that capture short-range dependencies.

This great success of the Self-Attention-based architectures encourages the community to directly re-use them in more constrained applications of these tasks.
Streaming Automatic Speech Recognition\cite{variani2022global} is an example of such a task, where models should begin to transcribe the speech input into text without having the full speech context. This means that models in Streaming ASR process speech input as chunks and do not take the full-context of speech while transcribing. 
Using small chunks while training the ASR models limits the visibility of the Self-Attention within these chunks. 
The result is a potential mismatch between architecture and task: the Self-Attention module, designed to exploit global context, is forced to operate locally while keeping its heavier cost profile.

For this specific task, many recent works bring improvements in terms of transcription quality and latency of the models\cite{yu2020dual,yu2021fastemit,kang2023delay,song2023trimtail,tsunoo2024decoder}, but the proposed improvements never question architectures that have been designed for standard task, i.e., non-streaming ASR.
\cite{parcollet2025summary} and \cite{moriya2025attention} show that replacing Self-Attention by a linear alternative that summarizes input information into a global vector or a Mamba module\cite{gu2023mamba} in Streaming ASR architectures preserves the performance of the system.
However, the reason why Self-Attention does not bring much improvement in Streaming ASR and why it could be replaced by alternative modules remains unexplored by the community, to the best of our knowledge.

This paper addresses these questions from the standpoint of encoder design for streaming ASR. We adopt a strict streaming setup in which models are trained and evaluated with fixed chunk sizes and no access to past context beyond the current chunk. Within this regime, we first analyse how Self-Attention actually behaves inside a Conformer encoder. Visualizing layer-wise mean attention maps with chunked inference reveals dominant near-diagonal patterns, indicating that attention is predominantly capturing short-range dependencies already well handled by the convolution module. This finding suggests that, under streaming constraints, self-attention effectively operates as an expensive local operator.

Guided by this analysis, we revisit the encoder architecture and study two pragmatic modifications that reduce cost while preserving accuracy. In a \textit{soft} approach, we substitute Self-Attention with a lightweight 1-D deformable convolution module, which adaptively focuses on local patterns within each chunk while maintaining the conformer’s overall structure. In a \textit{hard} approach, we remove Self-Attention, relying on the Conformer’s convolution module to capture local and chunk-level patterns. These designs are evaluated in a Conformer-Transducer framework on LibriSpeech and TEDLIUM-2 across multiple chunk sizes (160–1280 ms), using strictly streaming training/inference.

\begin{figure*}[!h]
    \centering
    \includegraphics[width=\textwidth]{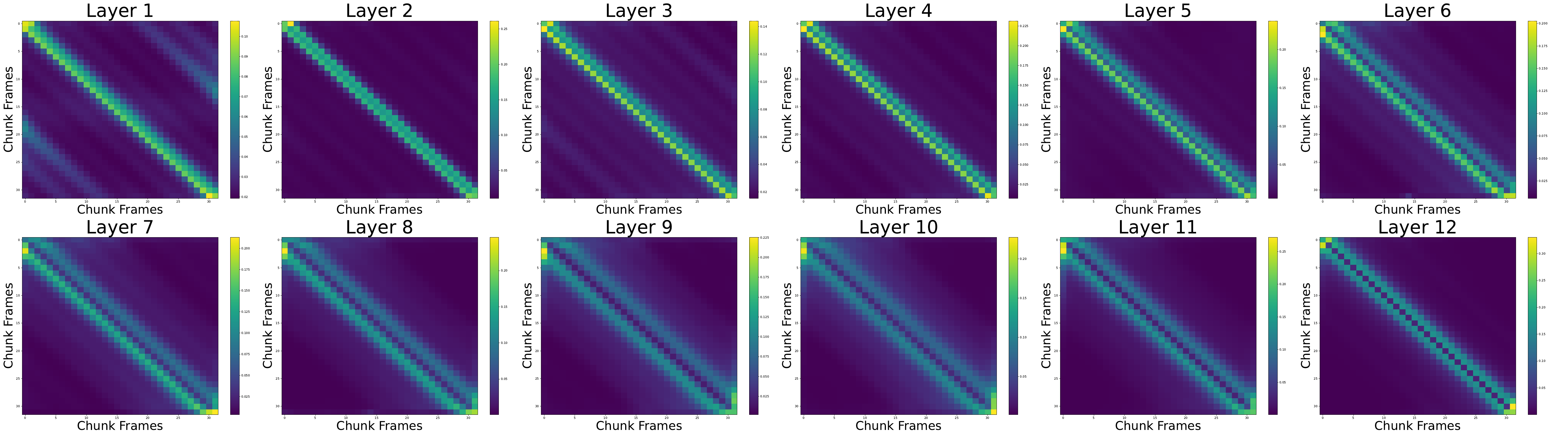}
    \caption{Illustration of the mean attention map per layer using chunk size 1280 ms on LibriSpeech test-clean dataset.}
    \label{fig:attention_map}
\end{figure*}

\section{preliminary analysis}

In this Section, we perform a preliminary analysis by examining how Self-Attention behaves in a strict streaming setting within a Conformer-Transducer architecture, which we consider as a baseline in the whole paper.
We choose the transducer architecture\cite{graves2012sequence} as it appears to be more suitable for streaming ASR as shown in \cite{kumar2025xlsr,li2021better,li2020towards}. We adopt a strict chunked regime at both training and inference: the model processes fixed-size chunks with no access to past or future context beyond the current chunk.

Figure \ref{fig:attention_map} shows the mean attention map calculated for each layer of a 12-layer conformer encoder in our baseline. We train and test this model using a chunk size of 1280 ms (32 audio frames). We see that, across all layers, attention concentrates in narrow bands around the main diagonal, with the strongest weights near the centre of the chunk. This pattern indicates that, in the streaming mode, Self-Attention predominantly captures short-range dependencies within each chunk. Given that the Conformer’s convolution module uses a kernel size of 31 and is stacked across layers, its effective receptive field within a 32-frame chunk can span most of the chunk. As a result, the convolution module likely aggregates chunk-level information, effectively reversing the canonical roles described for full-context Conformers, where Self-Attention models global dependencies and convolution focuses on local patterns.

To empirically confirm this hypothesis, we test, \textbf{without fine-tuning}, the baseline while masking all attention maps except their central diagonals. We obtain this configuration by masking the attention maps, before applying Softmax function, using the following mask $M$ where $N_{diag}$ is the number of the non-masked central diagonals.
\[
M_{i,j} = 
\begin{cases}
1, & \text{if } |i - j| \leq \left\lfloor \frac{N_{diag}}{2} \right\rfloor \\
0, & \text{otherwise}
\end{cases}
\]

\begin{table}
\centering
\caption{Word Error Rate (WER) measurements on a chunk size of 1280 ms with attention masking. The values in brackets correspond to the absolute difference between the result and the baseline with no masking.}
\label{tab:attention_masking}
\begin{tabular}{|c|c|c|}
\hline
\begin{tabular}[c]{@{}c@{}}Number of\\ non-masked diagonals\end{tabular} &
  \begin{tabular}[c]{@{}c@{}}LibriSpeech\\ test-clean\end{tabular} &
  \begin{tabular}[c]{@{}c@{}}LibriSpeech\\ test-other\end{tabular} \\ \hline
all & 3.36        & 8.91         \\ \hline
7          & 3.84(+0.48) & 9.96(+1.05)  \\ \hline
5          & 4.17(+0.81) & 10.81(+1.90) \\ \hline
\end{tabular}
\end{table}

Table \ref{tab:attention_masking} shows the results using this configuration while keeping only 7 or 5 non-masked diagonals in the attention maps, i.e., keeping, respectively, 21\% or 15\% of the values of each attention map. Despite the expected degradation, the model sustains reasonable performance, supporting the interpretation that, under strict streaming constraints, Self-Attention module operates primarily as an expensive local operator, while the convolution module carries much of the chunk-level information.

These observations suggest two pragmatic directions: replacing self-attention with a lighter local operator tuned for chunked inputs, or removing it altogether and relying on the convolution module to capture both local and chunk-level patterns. We explore both options in the next section.  

\section{Attention replacement}
Motivated by the preliminary analysis, we revisit the encoder design, in this Section, to show how much Self-Attention contributes under strict chunked streaming. 
We use the conformer encoder as a baseline of the model used in Streaming ASR task. This encoder is composed of a convolution subsampling block and a stack of conformer blocks. Each conformer block features a sandwich architecture, with a feed-forward module positioned at both the beginning and the end, enclosing a Self-Attention module and a convolutional module within. In our experiments, we explore two approaches to replace self-attention and evaluate its effectiveness. 

\noindent Note that we deliberately do not adopt efficient attention variants such as Fastconformer\cite{rekesh2023fast}, Linformer-style linear attention\cite{wang2020linformer}, local/blockwise attention, or memory-based streaming Transformers (Emformer\cite{shi2021emformer}, Zipformer\cite{yao2023zipformer}) for two reasons. First, our goal is to isolate the intrinsic contribution of self-attention under strict streaming constraints, where no past context or look-ahead is available. Many of these approaches explicitly rely on external memory/caches. Including them would obscure whether attention itself remains useful in this regime. Second, even efficient attention variants have non-trivial constant factors and implementation complexity that differ from standard attention and from convolutions; comparing them fairly would require substantial engineering and careful latency-quality trade-off tuning beyond the scope of this first study. Our objective here is to establish a clear, controlled baseline result: in strict chunked streaming, attention behaves primarily as a local operator and can be replaced by lightweight convolutional modules, or even removed, without significant loss in accuracy.

\subsection{Hard Approach}

In this approach, we completely remove the Self-Attention module without modifying the rest of the architecture. We suppose that the convolution module is solely capable of extracting all the diagonal patterns that could be captured by the Self-Attention inside the chunk due to the fact that its kernel size, 31 as in \cite{gulati2020conformer}, is larger than or equal to the chunk sizes tested in streaming inference.

\subsection{Soft Approach}

This approach consists of replacing the Self-Attention module with a deformable convolution module composed of a 1-D deformable convolution\cite{dai2017deformable} followed by a layer normalisation and a Swish activation layer.

\begin{figure}
    \centering
    \includegraphics[width=\linewidth]{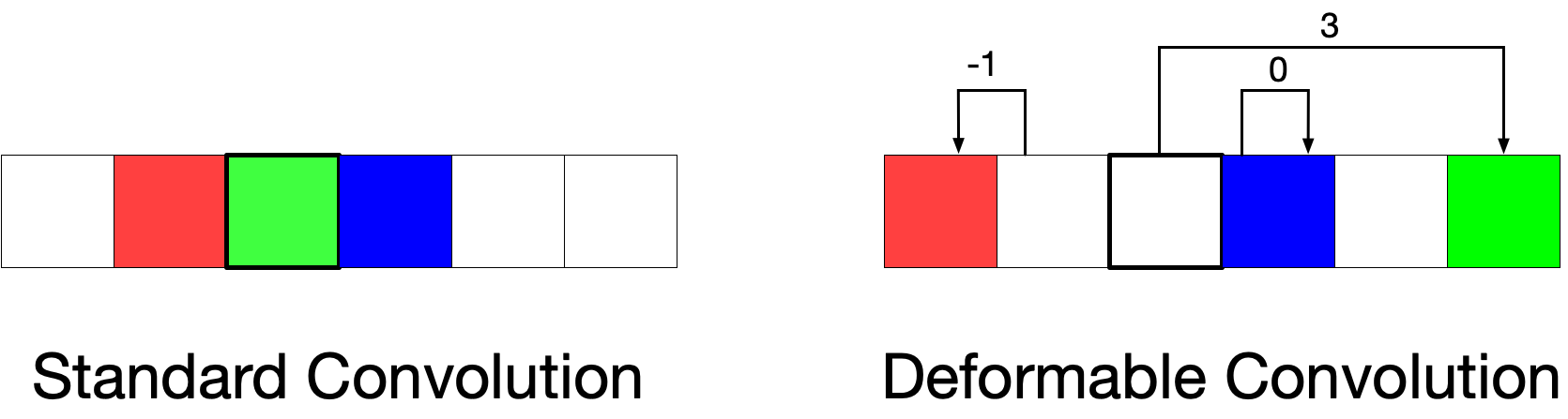}
    \caption{Illustration of a 1-D standard convolution (left) and 1-D deformable convolution (right) with kernel size of 3 applied at the $3^{rd}$ timestep. The coloured squares represent the elements on which the kernel of each convolution will be applied. The offsets applied to the deformable convolution in this example are [-1,3,0].}
    \label{fig:deformable_convolution}
    \vspace{-0.2cm}
\end{figure}

The deformable convolution, which is the core of our module, is a type of convolution with an asymmetric kernel. It works in 2 steps: first an offset convolution is applied on the input to predict input position offsets. Then, an output convolution is applied to the input taking into account the predicted input position offsets to shift its kernel elements, as shown in Figure \ref{fig:deformable_convolution}.

We choose using a deformable convolution rather than a standard convolution because the standard convolution processes all regions equally when a convolution kernel slides across the input audio where a deformable convolution can help build local patterns by focusing on the most relevant information.
\section{Experiments}
\subsection{Evaluation protocol}
We conduct our experiments on two widely used public datasets: LibriSpeech\cite{panayotov2015librispeech} that contains 960 hours of read English speech and TEDLIUM-2 \cite{zhou2020rwth} that contains 207 hours of TED Talks in order to show the effectiveness of our approach on more spontaneous speech.

To evaluate our approach on the Streaming-ASR task, we use the Word Error Rate metric. Also, to show the significance of the results, each result of our baseline is presented together with its confidence interval\footnote{https://github.com/luferrer/ConfidenceIntervals}. These intervals are calculated using the bootstrapping method with 1,000 bootstrap sets. They were also calculated between $2.5$ and $97.5$ percentiles to exclude outliers.

\vspace{-0.2cm}
\subsection{Model configuration}

All experiments 
were conducted using the SpeechBrain toolkit\cite{ravanelli2021speechbrain}. The baseline used for Streaming-ASR task and the proposed systems is Conformer Transducer which share the following components:
    \textbf{(a)} \textbf{2 Convolutional layers} with kernel size of 2 and stride of 2, which downsamples the frame rate by 4.
    \textbf{(b)} \textbf{12-layer Conformer encoder}\cite{gulati2020conformer} having 512-dimensional input where each layer is composed of: feed-forward network of size 2048, convolution block having kernel size of 31 with stride of 1 and a self-attention block with 8 attention heads in the case of the baseline. In the soft approach, we replace it with 1-D deformable convolution that has a kernel size of 5 and uses 8 groups.\footnote{https://github.com/inspiros/tvdcn/tree/master/tvdcn}
    \textbf{(c)} \textbf{Predictor network} of 1-layer LSTM\cite{graves2012long} with hidden size of 512.

During the training stage, we train our models for 150 epochs using transducer loss and an auxiliary CTC loss\cite{graves2006connectionist} for the first 10 epochs. In addition, an Adam optimizer is set with a learning rate of $0.0008$ and a weight decay of $0.01$. All our models are trained and tested using a unique chunk size without taking into account any past context.
\vspace{-0.2cm}
\subsection{Results}

\begin{table*}[!htbp]
\caption{the Word Error Rate measurements on the LibriSpeech test-clean, test-other and TEDLIUM-2 datasets with different chunk sizes. The values in brackets correspond to the relative reduction compared to the baseline and the values in square brackets represent the confidence interval of the result.}
\vspace{-0.2cm}
\centering
\begin{tabular}{|c|c|c|cccc|}
\cline{1-7}
\multirow{2}{*}{Dataset} &
  \multirow{2}{*}{Model} &
  \multirow{2}{*}{Number of parameters} &
  \multicolumn{4}{c|}{Chunk size} \\
 &
   &
   &
  160ms &
  320ms &
  640ms &
  1,280ms \\ \hline\hline
\multirow{3}{*}{\begin{tabular}[c]{@{}c@{}}LibriSpeech\\ test-clean\end{tabular}} &
  Baseline &
  81.3M &
  \multicolumn{1}{c|}{\begin{tabular}[c]{@{}c@{}}4.21\\ \scriptsize{{[}3.98;4.47{]}}\end{tabular}} &
  \multicolumn{1}{c|}{\begin{tabular}[c]{@{}c@{}}3.85\\ \scriptsize{{[}3.62;4.10{]}}\end{tabular}} &
  \multicolumn{1}{c|}{\begin{tabular}[c]{@{}c@{}}3.69\\ \scriptsize{{[}3.47;3.92{]}}\end{tabular}} &
   {\begin{tabular}[c]{@{}c@{}}3.36\\ \scriptsize{{[}3.15;3.58{]}}\end{tabular}} \\ \cline{2-7} 
 &
  Soft approach &
  67.6M(\textbf{-16.8\%}) &
  \multicolumn{1}{c|}{\begin{tabular}[c]{@{}c@{}}\textbf{4.11}\\
  \scriptsize{{[}3.90;4.35{]}}\end{tabular}} &
  \multicolumn{1}{c|}{\begin{tabular}[c]{@{}c@{}}3.86\\
  \scriptsize{{[}3.64;4.09{]}}\end{tabular}} &
  \multicolumn{1}{c|}{\begin{tabular}[c]{@{}c@{}}3.75\\
  \scriptsize{{[}3.54;3.98{]}}\end{tabular}} &
   \begin{tabular}[c]{@{}c@{}}3.56\\
   \scriptsize{{[}3.35;3.80{]}}\end{tabular}\\ \cline{2-7}
 &
  Hard approach &
  65.5M(\textbf{-19.4\%}) &
  \multicolumn{1}{c|}{\begin{tabular}[c]{@{}c@{}}4.29\\
  \scriptsize{{[}4.06;4.53{]}}\end{tabular}} &
  \multicolumn{1}{c|}{\begin{tabular}[c]{@{}c@{}}4.04\\
  \scriptsize{{[}3.81;4.29{]}}\end{tabular}} &
  \multicolumn{1}{c|}{\begin{tabular}[c]{@{}c@{}}3.78\\
  \scriptsize{{[}3.56;4.00{]}}\end{tabular}} &
   \begin{tabular}[c]{@{}c@{}}3.62\\
   \scriptsize{{[}3.41;3.84{]}}\end{tabular}\\ \hline\hline
\multirow{3}{*}{\begin{tabular}[c]{@{}c@{}}LibriSpeech\\ test-other\end{tabular}} &
  Baseline &
   81.3M &
  \multicolumn{1}{c|}{\begin{tabular}[c]{@{}c@{}}11.06\\ \scriptsize{{[}10.59;11.49{]}}\end{tabular}} &
  \multicolumn{1}{c|}{\begin{tabular}[c]{@{}c@{}}10.36\\ \scriptsize{{[}9.92;10.79{]}}\end{tabular}} &
  \multicolumn{1}{c|}{\begin{tabular}[c]{@{}c@{}}9.78\\ \scriptsize{{[}9.38;10.18{]}}\end{tabular}} &
   {\begin{tabular}[c]{@{}c@{}}8.91\\ \scriptsize{{[}8.52;9.28{]}}\end{tabular}}\\ \cline{2-7} 
 &
  Soft approach &
  67.6M(\textbf{-16.8\%}) &
  \multicolumn{1}{c|}{\begin{tabular}[c]{@{}c@{}}\textbf{11.03}\\
  \scriptsize{{[}10.62;11.43{]}}\end{tabular}} &
  \multicolumn{1}{c|}{\begin{tabular}[c]{@{}c@{}}\textbf{10.33}\\
  \scriptsize{{[}9.90;10.74{]}}\end{tabular}} &
  \multicolumn{1}{c|}{\begin{tabular}[c]{@{}c@{}}9.81\\
  \scriptsize{{[}9.37;10.23{]}}\end{tabular}} &
   \begin{tabular}[c]{@{}c@{}}9.34\\
   \scriptsize{{[}8.94;9.72{]}}\end{tabular}\\ \cline{2-7} 
 &
  Hard approach &
  65.5M(\textbf{-19.4\%}) &
  \multicolumn{1}{c|}{\begin{tabular}[c]{@{}c@{}}11.23\\
  \scriptsize{{[}10.78;11.64{]}}\end{tabular}} &
  \multicolumn{1}{c|}{\begin{tabular}[c]{@{}c@{}}10.39\\
  \scriptsize{{[}9.98;10.79{]}}\end{tabular}} &
  \multicolumn{1}{c|}{\begin{tabular}[c]{@{}c@{}}9.81\\
  \scriptsize{{[}9.42;10.21{]}}\end{tabular}} &
   \begin{tabular}[c]{@{}c@{}}9.62\\
   \scriptsize{{[}9.24;10.00{]}}\end{tabular}\\ \hline\hline
\multirow{3}{*}{TEDLIUM-2} &
  Baseline &
  81.3M &
  \multicolumn{1}{c|}{\begin{tabular}[c]{@{}c@{}}11.12\\ \scriptsize{{[}10.56;11.66{]}}\end{tabular}} &
  \multicolumn{1}{c|}{\begin{tabular}[c]{@{}c@{}}10.32\\ \scriptsize{{[}9.82;10.80{]}}\end{tabular}} &
  \multicolumn{1}{c|}{\begin{tabular}[c]{@{}c@{}}9.77\\ \scriptsize{{[}9.28;10.23{]}}\end{tabular}} &
   {\begin{tabular}[c]{@{}c@{}}9.26\\ \scriptsize{{[}8.79;9.73{]}}\end{tabular}} \\ \cline{2-7} 
 &
  Soft approach &
  67.6M(\textbf{-16.8\%}) &
  \multicolumn{1}{c|}{\begin{tabular}[c]{@{}c@{}}\textbf{11.04}\\
  \scriptsize{{[}10.52;11.60{]}}\end{tabular}} &
  \multicolumn{1}{c|}{\begin{tabular}[c]{@{}c@{}}10.49\\
  \scriptsize{{[}9.98;10.99{]}}\end{tabular}} &
  \multicolumn{1}{c|}{\begin{tabular}[c]{@{}c@{}}10.06\\
  \scriptsize{{[}9.55;10.55{]}}\end{tabular}} &
   \begin{tabular}[c]{@{}c@{}}9.27\\
   \scriptsize{{[}8.81;9.73{]}}\end{tabular}\\ \cline{2-7} 
 &
  Hard approach &
  65.5M(\textbf{-19.4\%}) &
  \multicolumn{1}{c|}{\begin{tabular}[c]{@{}c@{}}11.47\\
  \scriptsize{{[}10.93;11.99{]}}\end{tabular}} &
  \multicolumn{1}{c|}{\begin{tabular}[c]{@{}c@{}}11.08\\
  \scriptsize{{[}10.54;11.59{]}}\end{tabular}} &
  \multicolumn{1}{c|}{\begin{tabular}[c]{@{}c@{}}10.29\\
  \scriptsize{{[}9.81;10.81{]}}\end{tabular}} &
  \multicolumn{1}{c|}{\begin{tabular}[c]{@{}c@{}}9.87\\
  \scriptsize{{[}9.35;10.34{]}}\end{tabular}} \\ \cline{1-7}
\end{tabular}
\label{tab:results_all_approaches}
\end{table*}


\begin{figure}
    \centering
    \begin{subfigure}[t]{0.49\linewidth}
        \centering
        \includegraphics[width=\linewidth]{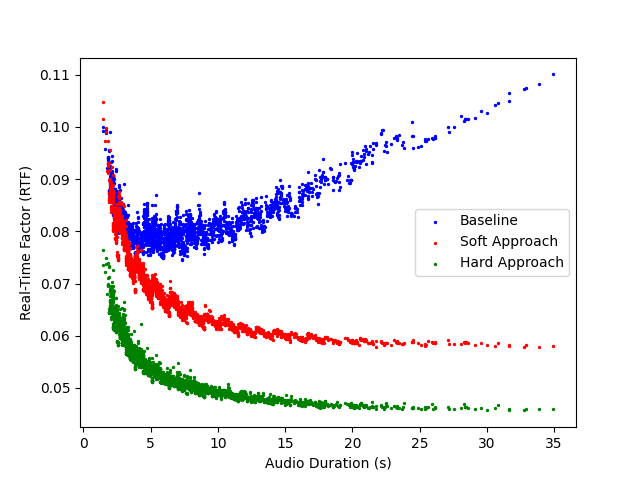}
        \caption{CPU}
        \label{subfig:cpu}
    \end{subfigure}
    \begin{subfigure}[t]{0.49\linewidth}
    \centering
        \includegraphics[width=\linewidth]{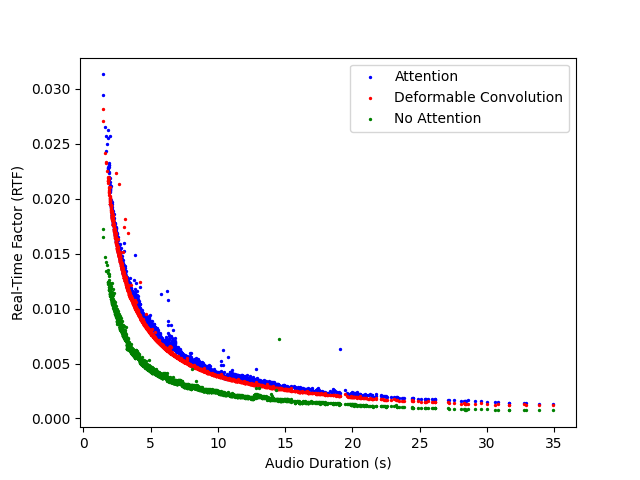}
        \caption{GPU}
        \label{subfig:gpu}
    \end{subfigure}
    \begin{subfigure}[t]{0.49\linewidth}
    \centering
        \includegraphics[width=\linewidth]{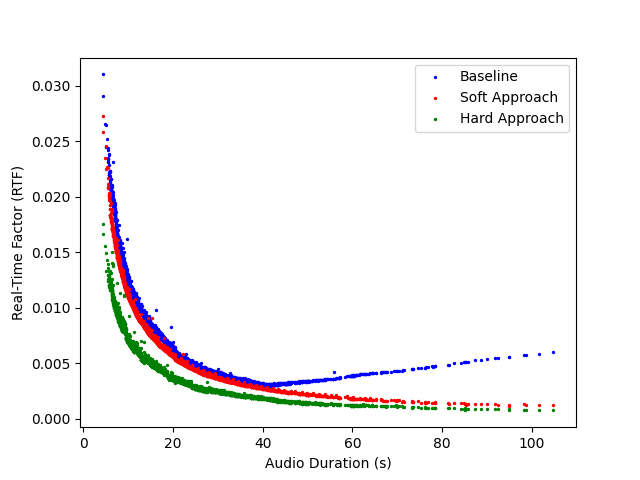}
        \caption{GPU on augmented dataset}
        \label{subfig:gpu_aug}
    \end{subfigure}
    \caption{Comparison of Real Time Factor (RTF) between the baseline, soft and hard approach for each utterance on LibriSpeech test-clean dataset. In \ref{subfig:gpu_aug}, the utterances have been repeated to be x3 longer.}
    \label{fig:rtf_models}
    \vspace{-0.2cm}
\end{figure}

Table \ref{tab:results_all_approaches} shows the Word Error Rate obtained by the different approaches compared to the baseline in different chunk sizes. We first see that using soft or hard approach implies an important reduction in terms of the number of parameters. Moreover, both approaches present insignificant degradation (in the sense of confidence intervals) compared to the baseline. The soft approach even outperforms the baseline while testing with small chunk sizes. This may be explained by the great capacity of deformable convolutions to capture local patterns compared to the Self-Attention module. 

To compare the efficiency of the proposed approaches in terms of computational cost, we measure the Real Time Factor (RTF) for each utterance of the LibriSpeech test-clean dataset, as we illustrate in Figure \ref{fig:rtf_models}. We use the same method as in \cite{parcollet2025summary}, by calculating the RTF on the encoder only, because the other components stay the same for all trained models. The measurements were made using the CPU \textit{AMD EPYC 7282-2.8GHz} and the GPU \textit{RTX 4070}. We observe that both soft and hard approaches reduce the computational cost of processing the utterances compared to the baseline. In fact, while using CPU (Figure \ref{subfig:cpu}),  the mean RTF across the dataset for the baseline is $8.18.10^{-2}$ where the soft approach gives an RTF of $6.89.10^{-2}$, which is 15.8\% faster. For the hard approach, it has an RTF of $4.51.10^{-2}$ which makes it 44.9\% faster than the baseline. For the GPU, the mean RTF across the dataset for the baseline is $8.21.10^{-3}$ where the soft approach gives an RTF of $7.78.10^{-3}$, which is 5.2\% faster. For the hard approach, it has an RTF of $4.51.10^{-3}$ which makes it approximately 2 times faster than the baseline.

On GPU, Figure \ref{subfig:gpu}, the quadratic scaling of self-attention with sequence length is largely masked by kernel-level parallelism and highly optimized Transformer implementations. To expose this behaviour, we extended the input by repeating each test utterance three times, as shown in Figure \ref{subfig:gpu_aug}. Under this long-utterance regime, the Self-Attention cost becomes evident, like in CPU, with the quadratic trend emerging for inputs longer than 45 seconds.  Consequently, for datasets with long recordings, the soft and hard approaches deliver larger RTF reductions than the baseline on GPU.

\vspace{-0.2cm}
\subsection{Ablation studies}
\begin{table}[]
\caption{Ablation study on the number of parameters using chunk sizes 320 ms and 640 ms. Performance are given in terms of WER and the values in brackets correspond to the absolute difference compared to the baseline in Table \ref{tab:results_all_approaches}.}
\vspace{-0.2cm}
\label{tab:ablation_num_params}
\resizebox{\linewidth}{!}{
\begin{tabular}{|c|c|c|c|c|}
\hline
Model &
  \begin{tabular}[c]{@{}c@{}}Number of \\ parameters\end{tabular} &
  \begin{tabular}[c]{@{}c@{}}Chunk \\ size\end{tabular} &
  \begin{tabular}[c]{@{}c@{}}LibriSpeech\\ test-clean\end{tabular} &
  \begin{tabular}[c]{@{}c@{}}LibriSpeech\\ test-other\end{tabular} \\ \hline
\multirow{3}{*}{\begin{tabular}[c]{@{}c@{}}Soft\\ approach\end{tabular}} & \multirow{3}{*}{79.7M} & 320 ms & 3.77\textbf{(-0.08)} & 10.22\textbf{(-0.14)} \\ \cline{3-5} 
& & 640 ms & 3.73(+0.04) & 9.62\textbf{(-0.16)} \\ \cline{3-5}
                               &                              & 1280 ms & 3.44(+0.08) & 9.15(+0.24)  \\ \hline
\multirow{3}{*}{\begin{tabular}[c]{@{}c@{}}Hard\\ approach\end{tabular}} & \multirow{3}{*}{79.6M} & 320 ms & 3.92(+0.07) & 10.44(+0.08) \\ \cline{3-5} 
& & 640 ms & 3.60\textbf{(-0.09)} & 9.84(+0.06) \\ \cline{3-5}
                               &                              & 1280 ms & 3.52(+0.16) & 9.18(+0.27)  \\ \hline
\end{tabular}}
\vspace{-0.2cm}
\end{table}

As shown in Table \ref{tab:ablation_num_params}, we conducted the first ablation studies to show the impact of the architecture compared to the number of parameters used in the model. For that, we train the soft and hard approaches using a number of parameters equivalent to the baseline. We achieve this by increasing the embedding dimension. We observe from Table \ref{tab:ablation_num_params} that the soft approach surpasses the baseline even when using larger chunk sizes. Moreover, the hard approach matches the baseline results for these large chunk sizes, which confirms that the convolution module alone is capable of capturing the local and global patterns inside the chunk. In conclusion, the insignificant degradations observed in Table \ref{tab:results_all_approaches} in larger chunk sizes are not related to the Self-Attention module but to the number of parameters.

\begin{table}[]
\centering
\caption{Ablation study on deformable convolution kernel sizes, used in the soft approach, trained on 1280 ms chunk size.}
\vspace{-0.2cm}
\label{tab:ablation_kernel_size}
\begin{tabular}{|c|c|c|c|}

\hline
Model & Kernel size & \begin{tabular}[c]{@{}c@{}}LibriSpeech\\ test-clean\end{tabular} & \begin{tabular}[c]{@{}c@{}}LibriSpeech\\ test-other\end{tabular} \\ \hline
\multirow{3}{*}{Soft approach} & 5  & 3.56        & 9.34        \\ \cline{2-4} 
                               & 9  & 3.54(-0.02) & 9.49(+0.15) \\ \cline{2-4} 
                               & 17 & 3.51(-0.05) & 9.04(-0.30) \\ \hline
\end{tabular}
\vspace{-0.2cm}
\end{table}

We conducted a second ablation study, described in Table \ref{tab:ablation_kernel_size}, to analyse the effect of the chosen kernel size for the deformable convolution module. We observe that augmenting the kernel size does not bring a significant improvement in LibriSpeech test-clean dataset and inconsistent results when testing on LibriSpeech test-other. These results confirm again that the deformable convolution needs to capture, in most cases, the short-range dependencies, and it does not require a wide kernel.  

\section{Conclusion}
Our study reveals that Self-Attention mechanisms in streaming ASR models may not be as essential as previously assumed, and found that it behaves as a local operator: attention maps concentrate near the diagonal, and constraining them to narrow central bands yields only minor WER changes. Guided by these findings, we proposed two encoder variants within a Conformer-Transducer, soft and hard approaches, that garde the performances or even surpass Attention-based model with significant computational cost and parameters reduction.
Future work should focus on developing architectures specifically designed for streaming ASR that prioritise local information processing. Given our findings, architectures that replace Self-Attention with more efficient modules like specialised convolutional modules, inspired by architectures such as Contextnet\cite{han2020contextnet} and Quartznet\cite{kriman2020quartznet}, should be explored.

\clearpage
\section{Acknowledgements}
This work was also granted access to the HPC resources of IDRIS under the allocation 2025-A0191014876 made by GENCI.
\bibliographystyle{IEEEbib}
\bibliography{refs}

\begin{thebibliography}{10}

\bibitem{vaswani2017attention}
A~Vaswani,
\newblock ``Attention is all you need,''
\newblock {\em Neurips}, 2017.

\bibitem{dosovitskiy2021an}
Alexey~Dosovitskiy et~al.,
\newblock ``An image is worth 16x16 words: Transformers for image recognition
  at scale,''
\newblock in {\em International Conference on Learning Representations}, 2021.

\bibitem{gulati2020conformer}
Anmol~Gulati et~al.,
\newblock ``Conformer: Convolution-augmented transformer for speech
  recognition,''
\newblock in {\em Interspeech 2020}, 2020, pp. 5036--5040.

\bibitem{variani2022global}
Ehsan~Variani et~al.,
\newblock ``Global normalization for streaming speech recognition in a modular
  framework,''
\newblock {\em Neurips}, vol. 35, pp. 4257--4269, 2022.

\bibitem{yu2020dual}
Jiahui~Yu et~al.,
\newblock ``Dual-mode asr: Unify and improve streaming asr with full-context
  modeling,''
\newblock in {\em International Conference on Learning Representations}, 2021.

\bibitem{yu2021fastemit}
Jiahui et~al. Yu,
\newblock ``Fastemit: Low-latency streaming asr with sequence-level emission
  regularization,''
\newblock in {\em ICASSP 2021-2021}. IEEE, 2021, pp. 6004--6008.

\bibitem{kang2023delay}
Wei~Kang et~al.,
\newblock ``Delay-penalized transducer for low-latency streaming asr,''
\newblock in {\em ICASSP 2023-2023}. IEEE, 2023, pp. 1--5.

\bibitem{song2023trimtail}
Xingchen~Song et~al.,
\newblock ``Trimtail: Low-latency streaming asr with simple but effective
  spectrogram-level length penalty,''
\newblock in {\em ICASSP 2023-2023}. IEEE, 2023, pp. 1--5.

\bibitem{tsunoo2024decoder}
Emiru~Tsunoo et~al.,
\newblock ``Decoder-only architecture for streaming end-to-end speech
  recognition,''
\newblock in {\em Interspeech 2024}, 2024, pp. 4463--4467.

\bibitem{parcollet2025summary}
Titouan~Parcollet et~al.,
\newblock ``Linear time complexity conformers with summarymixing for streaming
  speech recognition,''
\newblock in {\em ICASSP 2025 - 2025}, 2025, pp. 1--5.

\bibitem{moriya2025attention}
Takafumi~Moriya et~al.,
\newblock ``Attention-free dual-mode asr with latency-controlled selective
  state spaces,''
\newblock in {\em Proc. Interspeech 2025}, 2025, pp. 3588--3592.

\bibitem{gu2023mamba}
Albert Gu and Tri Dao,
\newblock ``Mamba: Linear-time sequence modeling with selective state spaces,''
\newblock {\em arXiv preprint arXiv:2312.00752}, 2023.

\bibitem{graves2012sequence}
Alex Graves,
\newblock ``Sequence transduction with recurrent neural networks,''
\newblock {\em arXiv preprint arXiv:1211.3711}, 2012.

\bibitem{kumar2025xlsr}
Shashi Kumar et~al.,
\newblock ``Xlsr-transducer: Streaming asr for self-supervised pretrained
  models,''
\newblock in {\em ICASSP}. IEEE, 2025.

\bibitem{li2021better}
Bo~Li et~al.,
\newblock ``A better and faster end-to-end model for streaming asr,''
\newblock in {\em ICASSP}. IEEE, 2021.

\bibitem{li2020towards}
Bo~Li, Shuo-yiin Chang, Tara~N Sainath, Ruoming Pang, Yanzhang He, Trevor
  Strohman, and Yonghui Wu,
\newblock ``Towards fast and accurate streaming end-to-end asr,''
\newblock in {\em ICASSP}, 2020.

\bibitem{rekesh2023fast}
Dima Rekesh et~al.,
\newblock ``Fast conformer with linearly scalable attention for efficient
  speech recognition,''
\newblock in {\em 2023 IEEE Automatic Speech Recognition and Understanding
  Workshop (ASRU)}. IEEE, 2023, pp. 1--8.

\bibitem{wang2020linformer}
Sinong Wang, Belinda~Z Li, Madian Khabsa, Han Fang, and Hao Ma,
\newblock ``Linformer: Self-attention with linear complexity,''
\newblock {\em arXiv preprint arXiv:2006.04768}, 2020.

\bibitem{shi2021emformer}
Yangyang Shi, Yongqiang Wang, Chunyang Wu, Ching-Feng Yeh, Julian Chan, Frank
  Zhang, Duc Le, and Mike Seltzer,
\newblock ``Emformer: Efficient memory transformer based acoustic model for low
  latency streaming speech recognition,''
\newblock in {\em ICASSP}. IEEE, 2021.

\bibitem{yao2023zipformer}
Zengwei Yao, Liyong Guo, Xiaoyu Yang, Wei Kang, Fangjun Kuang, Yifan Yang,
  Zengrui Jin, Long Lin, and Daniel Povey,
\newblock ``Zipformer: A faster and better encoder for automatic speech
  recognition,''
\newblock {\em CoRR}, 2023.

\bibitem{dai2017deformable}
Jifeng Dai, Haozhi Qi, Yuwen Xiong, Yi~Li, Guodong Zhang, Han Hu, and Yichen
  Wei,
\newblock ``Deformable convolutional networks,''
\newblock in {\em Proceedings of the IEEE international conference on computer
  vision}, 2017, pp. 764--773.

\bibitem{panayotov2015librispeech}
Vassil Panayotov, Guoguo Chen, Daniel Povey, and Sanjeev Khudanpur,
\newblock ``Librispeech: an asr corpus based on public domain audio books,''
\newblock in {\em 2015}. IEEE, 2015, pp. 5206--5210.

\bibitem{zhou2020rwth}
Wei Zhou, Wilfried Michel, Kazuki Irie, Markus Kitza, Ralf Schl{\"u}ter, and
  Hermann Ney,
\newblock ``The rwth asr system for ted-lium release 2: Improving hybrid hmm
  with specaugment,''
\newblock in {\em ICASSP}. IEEE, 2020.

\bibitem{ravanelli2021speechbrain}
Mirco Ravanelli, Titouan Parcollet, Peter Plantinga, Aku Rouhe, Samuele
  Cornell, Loren Lugosch, Cem Subakan, Nauman Dawalatabad, Abdelwahab Heba,
  Jianyuan Zhong, et~al.,
\newblock ``Speechbrain: A general-purpose speech toolkit,''
\newblock {\em arXiv preprint arXiv:2106.04624}, 2021.

\bibitem{graves2012long}
Alex Graves and Alex Graves,
\newblock ``Long short-term memory,''
\newblock {\em Supervised sequence labelling with recurrent neural networks},
  pp. 37--45, 2012.

\bibitem{graves2006connectionist}
Alex Graves, Santiago Fern{\'a}ndez, Faustino Gomez, and J{\"u}rgen
  Schmidhuber,
\newblock ``Connectionist temporal classification: labelling unsegmented
  sequence data with recurrent neural networks,''
\newblock in {\em ICML}, 2006.

\bibitem{han2020contextnet}
Wei Han, Zhengdong Zhang, Yu~Zhang, Jiahui Yu, Chung-Cheng Chiu, James Qin,
  Anmol Gulati, Ruoming Pang, and Yonghui Wu,
\newblock ``Contextnet: Improving convolutional neural networks for automatic
  speech recognition with global context,''
\newblock 2020.

\bibitem{kriman2020quartznet}
Samuel Kriman, Stanislav Beliaev, Boris Ginsburg, Jocelyn Huang, Oleksii
  Kuchaiev, Vitaly Lavrukhin, Ryan Leary, Jason Li, and Yang Zhang,
\newblock ``Quartznet: Deep automatic speech recognition with 1d time-channel
  separable convolutions,''
\newblock in {\em ICASSP}. IEEE, 2020.

\end{thebibliography}

\end{document}